\documentstyle[prl,aps,twocolumn]{revtex}

\tighten
\begin{document}
\draft
\title{Universal properties of aging in structural glasses}
\author{A. Latz }
\address{Technische Universit\"at Chemnitz, Institut f\"ur Physik,\\
  Reichenhainer Str. 70, D-09107 Chemnitz, Germany}

\date{June 5, 2001}
\maketitle
\begin{abstract} 
A microscopic theory 
for equilibrium and non equilibrium relaxations
in structural glasses is formulated. 
For all temperatures below the glass transition  
the dynamics can be asymptotically separated in a $\beta $ -
relaxation regime, which connects a quasi equilibrium with the aging
regime and the
aging regime. In the $\beta$ -  relaxation regime dynamic correlation
are independent from spatial correlations. 
This is due to a self organized critical
bifurcation scenario, which gives rise to one dynamic power
law and a logarithmic law.   
It is shown that
the fluctuation dissipation ratio in the aging regime is a universal
property of the system independent from  spatial scales.
\end{abstract}

\pacs{PACS numbers: 61.43.Fs, 61.20.Lc, 64.70.Pf, 05.70.Ln}

\section{Introduction}
Glasses are important solids, which are by definition
in an out of equilibrium state.  
Nevertheless in many experiments they
can be dealt with as if they were in a time independent {\em quasi
  equilibrium} state.
E.g. measurements of static structure factors and dynamic neutron or
light scattering functions give reproducible results seemingly 
independent of
the waiting time $t_w$ after the production of the glass. 
On the other side it is known, that properties of glasses depend on the
procedure with  which they are obtained and that they change on very
long time scales. The later phenomenon is known as aging of glasses
and there exists a vast experimental literature on this non
equilibrium behavior \cite{mckenna89}. 
So far it exists no microscopic theory of structural glasses,
which contains the equilibrium  and the non equilibrium aspects of
glasses, including  the influence of their non ergodic
character on the structure  and on the other side the influence
of the structure on the dynamics. 
\cite{foot1}. Although there are
many theoretical  investigations on the aging behavior of simple
models \cite{cugliandolo93,bouchaud98}, which are able to rationalize
properties of simulations on models of simple liquids out of equilibrium
\cite{kob97,barrat99,kob99,parisi97,parisi00},   they cannot, due to their
simplicity,  address
{\em structural} aspects of the aging
dynamics . For supercooled {\em liquids} in equilibrium a microscopic
theory, the mode 
coupling theory (MCT) of the glass transition, exists,  which
successfully describes the
dynamic of such systems at temperatures above the  critical  
glass transition
temperature \cite{leshouches,goetze99r}. The applicability of this
theory  below the transition temperature is not obvious, since it is based on
equilibrium considerations, which are not  fulfilled a priori for
non ergodic systems \cite{latz00b}. 
In this letter I will present a description of the dynamics
of glasses,  which unifies the non equilibrium aspects discussed in
simple model systems and the detailed structural aspects of the
(equilibrium) MCT of supercooled liquids. The  theory allows to draw
very general conclusions on the nature of aging in structural
glasses. 

The  goal is to develop  a theory, which allows to calculate dynamic
correlation and response functions in space and time for glasses. 
To describe the experimental situation, only the initial equilibrium
state is assumed to be known. The final glassy state is a result of
the dynamic  reached for time $t \to \infty$ after the quench at $t=0$. 
The variables relevant for the description of structural relaxation
are density fluctuations $n_q(t) = \sum_j \exp(i \vec{q} \vec{r}_j(t))$
and longitudinal current fluctuations $j_q(t) = \sum_j (\vec{q}
\vec{p_j})/(q m ) \exp(i \vec{q} \vec{r}_j(t))$. Here $\vec{r}_j $, 
$\vec{p_j}$ and $m$ are the position, the momentum  and the mass of
particle $j$. The density fluctuations contain the information about
the spatio temporal distributions of the particle, whereas the momentum
fluctuations are necessary to capture possible effects of global
momentum conservation and to describe the rate of the spatio temporal
changes. The equations of motion are given by the underlying
Hamiltonian dynamics and the coupling to a thermostat, which guarantees,
that the chosen thermodynamic parameters like density or temperature
are kept constant after the quench.  To be specific, I will choose in
the following a temperature quench from an initial equilibrium state
with temperature $T_i = 1/(\beta_i k_B)$ to a final temperature
$T$. The temperature can be measured with the help of the kinetic
energy $ k_B T = 2/(3 N) \langle \sum_j p_j^2/(2 m) \rangle
$, where $N$ is the number of particles. 
The average $\langle \ldots \rangle$ is an ensemble
average over the initial non equilibrium ensemble $\rho_{ne} = \exp
(-\beta \sum_j (p_j^2)/(2 m) - \beta_i \sum_{i \le j} V(|\vec{r}_i
- \vec{r}_j|))$, where $V(|\vec{r}_i
- \vec{r}_j|)$ is the interaction potential. 
This ensemble guarantees, that the value for
the temperature, measured with the help of the velocity distribution
is $k_B/\beta$ and that the spatial structure is the 
same as that of the liquid at the initial temperature
$k_B/\beta_i$. The dynamics is given by a time
dependent differential operator $L(t)$. 
For a variable $A(\vec{r}_i,
\vec{p}_i)$ the equation of motion is: $d A/d t = i U(t,0) L(t) A(0) 
:= U(t,0) \left( i L_0 A - 
\sum_j \alpha(t) \vec{p}_j(0) \partial A(\vec{r}_i,\vec{p}_i)/
\partial  \vec{p}_j(0) \right)$  
where  $U(t,0) = T_+ \exp(i \int_0^t L(t') d t')$ is a time ordered
product (time increases from left to right). $L_0 = -i \{H,.\}$ is
the Liouville operator, and $\{ ., . \}$ is the Poisson bracket.   
The Gaussian thermostat $\alpha(t)$ is adjusted such that $
\langle \sum_j (p_j^2(t))/(3 N m) \rangle = 1/\beta$  for
all times $t$. 
Given these equations of
  motion for the phase space variables,  one has to derive 
the equations of
  motion for the experimentally relevant correlation functions and,
  since a
  priori the validity of the FDT is not guaranteed,  
also for the susceptibilities.  A  suitable method to achieve this, is
a  non equilibrium projection  operator formalism, 
 using $n_q(t)$ and $j_q(t)$ as slow
  variables \cite{grabert82}.   
To obtain an {\em exact} equation for the
susceptibilities,  it is necessary to formulate the
projection operator in terms of Poisson brackets instead of
correlation functions \cite{latz01b}. 
 With such a projection operator and the specified iso--kinetic
dynamics the equation for
susceptibility $\chi_q(t_w,t) = \Theta(t-t_w)
[n_q(t_w)|n_q(t)]/N:=\langle \{n^*_q(t_w),n_q(t)\} \rangle /N $  and
the correlation function  $C_q(t_w,t) = (n_q(t_w)|n_q(t))/N :=
\langle n^*_q(t_w)|n_q(t)\rangle /N $, where
$t_w$ is the waiting time after the quench,  can be
derived as \cite{latz01b}

\begin{eqnarray} \label{sus}
\lefteqn{\frac{\partial ^2}{\partial t^2} \chi_q(t_w,t) =  -i \frac{m}{N}
\chi_q(t_w,t)  
[j_q(t)| L(t) j_q(t)]}\\
& -&  i m \, \alpha(t) \frac{\partial }{\partial t}
\chi_q(t_w,t) 
+  m \int_{t_w}^t d t' \chi_q(t_w,t')
\Sigma_q(t',t) \nonumber.
\end{eqnarray}

The memory function $\Sigma_q(t',t)$ is the average of the Poisson
bracket of the true force at $t=t'$ and the random force $F_q(t',t)$
between $t'$, 
and $t$. This random force describes the dissipation of the gradient of the
longitudinal stress at $t'$ in
non--hydrodynamic modes between $t'$ and $t$.  
The correlation function  $C_q(t_w,t)$ depends on all times
between $0$ and $t$, since the time translational invariance is
broken. The equation for it reads

\begin{eqnarray} \label{cor}
\lefteqn{\frac{\partial ^2}{\partial t^2} C_q(t_w,t) =  -i \frac{m}{N}
C_q(t_w,t) 
[j_q(t)| L(t) j_q(t)]}\nonumber \\
 &-&  i m \, \alpha(t) \frac{\partial }{\partial t}
C_q(t_w,t)  
 +  m \int_{0}^t d t' C_q(t_w,t')
\Sigma_q(t',t) \nonumber \\ 
&+& m \int_{0}^{t_w} d t' \chi_q(t',t_w)
M_q(t',t)\nonumber \\
&-& q \, C_q(t_w,0) \frac{1}{ N S_q(0)} (n_q(0)| F_q(0,t)) \\
&-& q \, (n_q(t_w)|j_q(0)) \frac{1}{N (j_q(0)|j_q(0))} (j_q(0)| F_q(0,t)
\nonumber     
\end{eqnarray}
The last two lines in Eq. (\ref{cor}) describe the correlation to the
initial state.  The memory function $M_q(t_w,t)$ is an autocorrelation
function of  random forces $F_q(0,t_w)$ and $F_q(0,t)$.  
The equation for the time dependent equal time correlation function
$S_q(t) = (n_q(t)|n_q(t))/N$, which describes the change in  the structure
of the liquid after the quench, is easily derived. 
\begin{equation} \label{struc}
\frac{1}{2}\frac{d^2}{dt^2} S_q(t) = (j_q(t)|j_q(t))/N - \frac{\partial
  ^2}{\partial t^2} C_q(t_w,t)|_{t_w=t}.
\end{equation}
 
If we make the assumption, that the velocities of different particles
are as in equilibrium uncorrelated at every time $t$, equation
(\ref{struc}) simplifies further, since $(j_q(t)|j_q(t))/N = 1/(\beta
m)$ in this case.

The set of equation (\ref{sus}-\ref{struc}) are  
formally exact.   If the initial state is an equilibrium state i.e. if
there is no temperature quench, it can be exactly shown \cite{latz01b},
that the correlation function and susceptibilities fulfill the FDT
\begin{equation} \label{FDT}
\chi_q(t_w,t) = \chi^{eq}_q(t-t_w) = - (1/T) d
  C_q^{eq}(t-t_w)/d t
\end{equation} 
 and the memory functions $\Sigma_q(t_w,t)$
and $M_q(t_w,t)$ obey a FDT of the second kind 
\begin{equation} \label{FDT2}
\Sigma_q(t_w,t) = \Sigma^{eq}_q(t-t_w) = - (1/T) d
  M_q^{eq}(t-t_w)/d t
\end{equation} 
  
The three equations (\ref{sus}, \ref{cor}) and (\ref{struc}) are
reducing then to one equation for the correlation function
\cite{latz01b}, 
which was the basis for the equilibrium MCT \cite{bengtzelius84}.  

A solvable theory is obtained by approximating the
two unknown functions $\Sigma_q(t_w,t)$ and $M_q(t_w,t)$. 
The simplest non trivial approximation consistent
with the mathematical requirement for the memory functions
\cite{latz01b} is a one loop
type or mode coupling approximation. It has
the form 
\begin{eqnarray} \label{sigmamct} 
\lefteqn{\Sigma_q(t_w,t) = q^2/(m^2 \beta^2)
\sum_{\vec{k},\vec{p}} \delta_{\vec{q},\vec{p}+\vec{k}} } \\
&&  c_{qkp}(\{S_q(t_w)\}) \;
c_{qkp}(\{S_q(t)\}) \chi_k(t_w,t)  C_p(t_w,t) \nonumber  
\end{eqnarray}
\begin{eqnarray} 
\lefteqn{M_q(t_w,t) = q^2/(2 m^2 \beta^2)
\sum_{\vec{k},\vec{p}} \delta_{\vec{q},\vec{p}+\vec{k}} } \label{mmct} \\
&& 
c_{qkp}(\{S_q(t_w)\}) \; c_{qkp}(\{S_q(t)\}) C_k(t_w,t)
C_p(t_w,t) \nonumber   
\end{eqnarray}

The vertices $c_{qkp}$ are functionals of the one time correlation
function $S_q(t)$.    They are symmetric with respect to exchange of
$p$ and $k$. Under this general condition, the FDT of the second kind
(\ref{FDT2})  is automatically fulfilled in equilibrium.  The   explicit
form of $c_{qkp}(t)$ will determine the theory for the structure factor
$S_q(t)$.   
For asymptotically long times $t_w,t \to \infty$ all correlation
functions, which depend 
only on a single time can be replaced by their asymptotic value.  If
the quench was performed from an equilibrium liquid state, the
functions $C_q(t_w,0)$  and 
$(n_q(t_w)|j_q(0))$ in Eq. (\ref{cor}) will vanish for $t_w \to \infty$.  It
can also be shown \cite{latz01b}, that the influence of the thermostat
$\alpha(t)$ 
vanishes at least as $1/t$ and therefore  the equations for the correlation
functions 
and the susceptibilities in the long time limit depend only on
the asymptotic value of the structure factor $S_q^\infty$ and the
vertices $c_{qkp} (\infty)$.

 If  there were no wave vector dependencies, the structure of
the resulting equations
were similar to the equations for the spherical $p$  - spin model
\cite{crisanti93,cugliandolo93} in the
same asymptotic limit. As in this model I will make the ansatz, that 
there exist below the glass transition two different dynamic regimes.  The
FDT regime  is characterized by the limit $C_q^{FDT}(\tau) =
\lim_{t_w\to \infty} C_q(t_w,t_w+\tau)$  with $\chi_q^{FDT}(\tau) =
(-1/T) d C_q^{FDT}(\tau)/d \tau$. The  aging regime is defined by the
limit  $\hat{C}_q(\lambda) = \lim_{t\to \infty}
C_q(\lambda t,t)$. This scaling behavior is called simple aging. 
There are more  general scenarios which could so far only be ruled out
for the $p$ -  spin model \cite{kim00}. For notational simplicity I will
concentrate on the ansatz of simple aging. 
I also will make the ansatz of a
generalized FDT ansatz, known to
be correct for the $p$ - spin model \cite{cugliandolo93}. It assumes, that
the susceptibility  $\hat{\chi}_q(\lambda) =
\lim_{t\to \infty} t \chi_q(\lambda t,t)$ is related to the
correlation function by
$\hat{\chi}_q(\lambda) = (X_q/T) 
d \hat{C}_q(\lambda)/d \lambda$ \cite{cugliandolo97}. Astonishingly it
turns out 
\cite{latz01b} that only a wave vector
independent fluctuation dissipation ratio (FDR) $ 0 \le X(T) \le 1$  
will give compatible equations for correlation
functions and susceptibilities in the aging regime.

With the definitions of normalized quantities  
$\phi_q(t_w,t) =
C_q(t_w,t)/S_q^\infty$,  $N_q(t_w,t) =
\chi_q(t_w,t)/S_q^\infty$, $f_q = F_q/S_q^\infty$, where
$F_q=\lim_{\tau \to \infty} C_q^{FDT}(\tau) = \hat{C}_q(1)$ and  $v_{qkp} =
\delta_{\vec{q},\vec{k} + \vec{p}} c^2_{qkp}(\infty) S_k^\infty
S_p^\infty$,   
the following equations for the
asymptotic structure factor $S_q^\infty$ is obtained 
\begin{equation}\label{struc3}
1/S_q^\infty =  m \beta c_B^2  - 1/2 
{\sum_{k,p}} 
v_{qkp} \left\{ \frac{}{} 1 -  
f_q f_k f_p (1- X)\right \}
\end{equation}
Here $c_B^2  = \lim_{t \to \infty} i m [j_q(t)|L_0 j_q(t)]/q^2$ 
is a bare wave vector dependent velocity, 
which will be renormalized by
contributions from the vertices $v_{qkp}$.
 The
equation for the correlation function in the FDT regime i.e.  $ -\beta
\; d
\phi_q^{FDT}(\tau)/d \tau = \chi_q^{FDT}(\tau)$ is given by 
\begin{eqnarray} \label{cor_fdt}
\lefteqn{ \frac{\partial ^2}{\partial \tau^2} \phi^{FDT}_q(\tau) =
-\left[c_B^2 
   q^2 - \frac{q^2}{2m\beta} \sum_{\vec{k} \vec{p}} v_{qkp}\right]   
\phi^{FDT}_q(\tau)} \\
&-& \frac{q^2}{2 m \beta} {\sum_{k,p}} v_{qkp} \left\{  
\int_{0}^\tau d \tau' \frac{ d \phi^{FDT}_q(\tau')}{ d \tau'}
(\phi^{FDT}_k \times \phi^{FDT}_p)(\tau-\tau') \nonumber \right.\\
 &+& \left. f_q f_k f_p
(1-X )  \right\}
\nonumber 
\end{eqnarray}

With Eq. (\ref{cor_fdt}) and (\ref{struc3}) it follows immediately
that Eq.  (\ref{cor_fdt}) reduces to the
standard MCT equation for the supercooled liquid   
 above the glass transition temperature
$T_c$, where $f_q=0$ and $X_q =0$.    It is interesting to
note, that like in the approach of Zaccarelli et
al. \cite{zaccarelli01b}, the mode coupling approach contains
implicitly a theory for the structure factor of liquids.   Below the glass
transition the structure of the glass depends on the non ergodicity
parameter i.e. the non ergodic nature of the glass also influences the equal
time correlation function.

\begin{eqnarray}\label{sus_aging}
0 &=&  c_B^2  + 1/(2 m \beta) \sum_{k,p} v_{qkp} \left\{(1 - 
f_k f_p)) \hat{N}_q(\lambda)) \right.\\ 
&+&  \hat{N}_k(\lambda)
\hat{\phi}_p(\lambda) (1 - f_q) 
+ \left. 1/ \beta \;    \int_\lambda^1
\frac{ds}{s} 
\hat{N}_q((\frac{\lambda}{s}))  \hat{N}_k(s)
\hat{\phi}_p(s)\right\}\nonumber  
\end{eqnarray}

 The
Eq. for the non ergodicity parameter can be obtained
from the $t \to \infty$ limit of (\ref{cor_fdt}) and it is identical
with the 
initial value $\hat{\phi}_q(1)$. With the help
of Eq. (\ref{struc3})  it can be written
\begin{equation}\label{nep}
f_q/(1-f_q) = \sum_{k,p} v_{qkp}  f_k f_p (1 -  (1-X) (1 -f_q))
\end{equation}

Since it can be shown, that $X=1$ for $T=T_c$ only, Eq. (\ref{nep}) reduces
to the equation of the MCT of supercooled liquids only at $T=T_c$. For
$T < T_c$ it is modified due to the breaking of the FDT ($X \ne 1$). 
 
A deeper insight in the meaning of the wave vector independence of $X$ can be
obtained by investigating the behavior of $\phi_q(t_w,t)$ for times,
where it is close to the non ergodicity parameter $F_q$. 
A detailed analysis shows \cite{latz01b},
that the Eqs. (\ref{cor_fdt}) and (\ref{sus_aging}) can be
simultaneously solved by the ansatz  $\phi_q(t_w,t) = f_q +
 e_q G(t_w,t)$ up to quadratic order in $G(t_w,t)$. Here $e_q$ is an
eigenvector of a matrix $D_{qk}$ 
\begin{eqnarray} 
0 &=&  \sum_k D_{qk'} {\hat{G}}'_{k'}(1)  \quad \mbox{with} \label{solve}\\
D_{qk'} &=& - (c_B^2 q^2  - (q^2/ 2 m \beta)
 \sum_{kp} v_{qkp} (1 - 
f_k f_p)) \delta_{q,k'}    \nonumber \\ 
&+& (q^2/2 m \beta) v_{qk'p} 
f_p (1 - f_q).   \label{dqk}
\end{eqnarray}
Eq. (\ref{solve}) is the solvability condition for
$\hat{N}_q(\lambda) = (X/T)  
d \hat{G}_q(\lambda)/d \lambda$  
by setting $\lambda =1$ in Eq. (\ref{sus_aging}).  

 The equations
for $G(t_w,t)$ in the FDT and the aging regime follow by multiplying
the resulting
equations from left with the left eigenvector $\hat{e}_q$ to
eigenvalue zero of $D_{qk'}$ i.e. $\sum_q \hat{e}_q D_{qk'} = 0$:
\begin{equation} \label{g_fdt}
0 = \Lambda \; \; (G^{FDT})^2(\tau) - \frac{d}{d \tau} \int_0^\tau d \tau'
G^{FDT}(\tau-\tau') G^{FDT}(\tau')
\end{equation}

\begin{equation}\label{g_aging}
0 = (\Lambda/X) {\hat{G}}^2(\lambda) - 
\int_\lambda^1 ds 
\frac{d \hat{G}(\frac{\lambda}{s})}{d s} \; \hat{G}(s)
\end{equation}

where 
\begin{equation} \label{l_parameter}
\Lambda = \frac{\sum_{qkp} \hat{e}_q \frac{q^2}{2 m \beta} v_{qkp}
  (1- f_q) e_k e_p}{\sum_{qkp} \hat{e}_q \frac{q^2}{ m \beta} v_{qkp}
   e_q e_k f_p}
\end{equation}
The microscopic structure of the system only enters via the exponent
parameter $\Lambda$, which is a complicated functional of the
asymptotic structure factor $S_q^\infty$. This behavior is very
reminiscent of the $\beta$ -  relaxation phenomena in supercooled liquids
above the glass transition.  
Eq. (\ref{g_fdt}) has exactly the form of the $\beta$  - scaling equation of
the MCT of supercooled liquids {\em at the critical temperature}
\cite{goetze85}.   In
this theory the glass transition is a critical bifurcation of codimension 1,
in which the  dynamics close to the transition temperature and for
times, 
at which the correlation function is  close to the non
ergodicity parameter, is completely determined by the critical direction
of the bifurcation. 
In the theory presented here, 
the additional  solvability condition for aging Eq. (\ref{solve})
forces the system
always to be at the critical bifurcation for temperatures below $T_c$.
Since this bifurcation is
still of codimension 1, there is only one temperature dependent number
to guarantee this {\em self organized criticality}, namely the
fluctuation dissipation ratio $X(T)$. This is the deeper reason, why
$X$ does not depend on wave vectors.  Within the presented theory,  aging
is due to a self organized critical bifurcation scenario. Because of
this criticality Eq. (\ref{g_fdt}) and (\ref{g_aging}) can be 
solved exactly for all temperatures below the transition temperature 
$G^{FDT}(\tau) = \tau^{-a} \quad \mbox{for} \;  t^* \gg \tau \gg
\tau_0 $  
where $\tau_0$ is some microscopic correlation scale and $t^*(t_w)$ is
the time where $\phi_q(t_w,t)= f_q$ \cite{foot2}. The exponent a
is determined by $\Lambda$:
\begin{equation} \label{a}
\Lambda(T) = \Gamma^2(1-a)/\Gamma(1-2a),
\end{equation}
where $\Gamma$ is the Euler Gamma function. 
The form of this equation is known from the $\beta$ -  scaling theory
of supercooled liquids \cite{goetze85}. But contrary to this theory
Eq. (\ref{a}) holds 
for {\em all} temperatures below the glass transition  with a temperature
dependent exponent parameter $\Lambda$ and thus a temperature
dependent exponent $a$. Equation (\ref{g_aging}) is
solved by 
$\hat{G}(\lambda) = -(-\log \lambda)^b$  for $\lambda <1$. 
 Here 
 $t_\alpha(t_w)$ is a 
 time characterizing the final decay of the correlation function 
  The exponent $b$  is
determined by an   equation similar to (\ref{a})
\begin{equation} \label{b}
\Lambda(T) = X(T) \; \Gamma^2(1+b)/\Gamma(1+2b)
\end{equation}
At the transition Eq. (\ref{b}) reduces to the Equation for the von
Schweidler exponent of the $\beta$ scaling theory \cite{goetze85}. 
Below the transition relation (\ref{b})  was also found in mean field theories
\cite{cugliandolo96a},  in
which the dynamics reduces for all times to a single component
theory. For models, in which simple aging does not hold,
$\lambda=t_w/t$ has to be replaced by $\lambda' = h(t_w)/h(t)$, where
$h(t)$ is a yet unspecified function (see e.g. the discussion in
\cite{kim00}).    
It is important to notice, that the decoupling of wave vector
dependence and time dependence in the present theory does not apply
once the condition ${\hat{G}}^3(\lambda) \ll \hat{G}^2(\lambda)$
is violated.  
In this case wave vector dependencies of the correlation functions
are to be expected and the solution of (\ref{sus_aging}) can only be
found numerically ($X(T)$ of course remains wave vector independent).  
The predicted factorization property $\phi_q(t_w,t) = f_q + e_q
G(t_w,t)$ was indeed found in simulations \cite{kob99}. The presented
theory allows for a systematic theoretical investigation of the
dynamics of the glassy
state of simple liquids as e.g. the colloidal hard sphere system, in
which indication for aging was already observed \cite{vanmegen98} or
the Lennard Jones systems of \cite{kob97,parisi00}.
 Especially it the foundation for a
theory of the structure factor of glasses by specifying the vertices
$c_{qkp}(t)$ in Eqs. (\ref{sigmamct}) and (\ref{mmct}). This will be
part of future work.      
\vspace{0.5cm}

{\bf  Acknowledgment}
It is a great pleasure to thank R. Schilling for his continuous support
and encouragement during the development of the presented theory. 
Financial support came
from the SFB262 of the Deutsche Forschungsgemeinschaft and the
Institut f\"ur Physik of the Johannes Gutenberg Universit\"at, Mainz.

\begin{thebibliography}{10}
\bibitem{foot1}The theory, 
which was formulated in \cite{latz00b},
although formally exact, contains too many unknown functions and is
therefore not suitable as starting point for real
calculations
\bibitem{foot2}The exact scaling of $t^*$ with
  $t_w$ is not known analytically. Numerical calculation and
  analytical estimates  for simple models \cite{kim00,zippold99}
  indicate, that $t^* \propto t_w^\gamma$ with 
  $\gamma < 1$
\end{thebibliography}

\end{document}